\documentstyle[12pt]{article}
\begin{document}
\begin{center}
LAG TIMES AND PARAMETER MISMATCHES IN  SYNCHRONIZATION OF  UNIDIRECTIONALLY COUPLED CHAOTIC EXTERNAL CAVITY SEMICONDUCTOR LASERS\\
E. M. Shahverdiev \footnote{Permanent address: Institute of Physics, 
370143 Baku, Azerbaijan}, S.Sivaprakasam and K. A. Shore \footnote{Electronic address: alan@sees.bangor.ac.uk}\\
School of Informatics, University of Wales, Bangor, Dean Street, Bangor, LL57 1UT, Wales, UK\\
ABSTRACT\\
\end{center}
We report an analysis of synchronization between two unidirectionally coupled chaotic external cavity master/slave semiconductor lasers with two characteristic delay times, where the delay time in the coupling is different from the delay time in the coupled systems themselves. We demonstrate for the first time that parameter mismatches in photon decay rates for the master and slave lasers can explain the experimental observation that the lag time is equal to the coupling delay time.\\
PACS number(s):05.45.Xt, 05.45.Vx, 42.55.Px, 42.65.Sf\\
\indent Chaos synchronization [1] is of fundamental importance in a variety of complex physical, chemical and biological systems [2]. Application of chaos synchronization has been advanced in secure communications, optimization of non-linear systems' performance, modeling brain activity and pattern recognition [2]. Time-delay systems are ubiquitous in nature, technology and society because of finite signal transmission times, switching speeds and memory effects [3]. Therefore the study of chaos synchronization in these systems is of considerable practical 
significance. Because of their ability to generate high-dimensional chaos, time-delay systems are good candidates for secure communications based on chaos synchronization. In this context particular emphasis is given to the use of chaotic external cavity 
semiconductor lasers, because laser systems with optical feedback are prominent representatives of time-delay systems which can generate hyperchaos [4]. \\
\indent Most experimental investigations of chaos synchronization in unidirectionally coupled external cavity semiconductor lasers [4] have found that the lag time between the master and slave lasers' intensities is equal to the coupling delay, whereas 
numerical results [5] show that the lag time should be equal to the difference between the delay time in the coupling and round-trip time of the light in the transmitter's external cavity. Knowledge of the exact lag time is of considerable practical importance, as experiments on message transmission using fibre lasers and diode lasers have shown that the recovery of message at the receiver critically depends on the correction made for the lag time [4,6].\\
\indent Recently there have been several attempts to explain the coupling-delay lag time synchronization in unidirectionally coupled external cavity semiconductor lasers.
In [7] this phenomenon was related to a strong coupling and/or frequency detuning between the two lasers. However in a recent paper [8], where a {\it numerical} study of two unidirectionally coupled single-mode semiconductor lasers subject to  optical 
feedback is reported, it was shown that such a phenomenon can be observed without any frequency detuning between the two lasers. In [8] it was found that two fundamentally different types of chaotic synchronization can occur depending on the relation between the  strengths of the coupling and of the feedback of the lasers. In the first type of synchronization, when the feedback rates of the transmitter and receiver lasers are equal, the lag time is equal to the coupling delay between the transmitter and receiver lasers; in the second type of synchronization, when the feedback rate of the transmitter is equal to the sum of the feedback rate of the receiver and coupling strength, the lag time is the difference between the coupling delay and the round-trip time of the light in the transmitter. In numerical investigations of the first type of synchronization  reported in [8] it was found that the synchronization error does not decay to zero but rather shows small oscillations even when the authors consider modified synchronizarion manifolds for the electric field amplitude and the carrier density by the introduction of a constant correction coefficient . Thus as the authors of [8] themselves acknowledge, the synchronization manifold introduced in [8] is not perfect but is approximate in nature.\\
\indent In this paper we derive existence conditions for two types of perfect synchronization between unidirectionally coupled master external cavity lasers:1)for the first time we demonstrate that parameter mismatches in photon decay rates for the master and slave lasers can explain the experimental observation that the lag time is equal to the coupling delay time. We compare our analytical findings with numerical 
simulations in [8] and discuss why the first type of synchronization numerically investigated in [8] was not perfect; 2)we recover the well-known result that 
for identical lasers the lag time in synchronization of master and slave systems is the difference between the coupling delay time and the time delay in the coupled systems themselves .\\
An appropriate framework for treating the evolution of the electric field of external cavity laser diodes is provided by the widely utilised Lang-Kobayashi equations [9]. Suppose that the master laser described by equations 
$$\hspace*{-1.7cm}\frac{dE_{1}}{dt}=\frac{(1+\imath\alpha_{1})}{2}(\frac{G_{1}(N_{1}-N_{01})}{1+s_{1}\vert E_{1} \vert^{2}}-\gamma_{1})E_{1}(t)+k_{1}E_{1}(t-\tau_{1})\exp(-\imath\omega\tau_{1}),$$
$$\hspace*{6cm}\frac{dN_{1}}{dt}=J_{1}-\gamma_{e1} N_{1}-\frac{G_{1}(N_{1}-N_{01})}{1+s_{1}\vert E_{1} \vert^{2}}\vert E_{1} \vert^{2},\hspace*{3.4cm}(1)$$
is coupled unidirectionally with the slave laser described by equations 
$$\hspace*{0.2cm}\frac{dE_{2}}{dt}=\frac{(1+\imath\alpha_{2})}{2}(\frac{G_{2}(N_{2}-N_{02})}{1+s_{2}\vert E_{2} \vert^{2}}-\gamma_{2})E_{2}(t)+k_{2}E_{2}(t-\tau_{1})\exp(-\imath\omega\tau_{1})+k_{3}E_{1}(t-\tau_{2})\exp(-\imath\omega\tau_{2}),$$
$$\hspace*{6cm}\frac{dN_{2}}{dt}=J_{2}-\gamma_{e2} N_{2}-\frac{G_{2}(N_{2}-N_{02})}{1+s_{2}\vert E_{2} \vert^{2}}\vert E_{2} \vert^{2},\hspace*{3.4cm}(2)$$
where $E_{1,2}$ are the slowly varying complex fields for the master and slave lasers,respectively;$N_{1,2}$ are the carrier densities;$\gamma_{1,2}$ are the cavity losses;$\alpha_{1,2}$ are the linewidth enhancement factors;$G_{1,2}$ are the optical gains;$k_{1,2}$ are the feedback levels;$k_{3}$ is the coupling rate;$\omega$ is the optical frequency without feedback (no frequency detuning between the two lasers);$\tau_{1}$ is the round-trip time in the external cavity;$\tau_{2}$ is the time of flight between the master laser 
and the slave laser-coupling delay time;$J_{1,2}$ are the injection currents;$\gamma_{e1,e2}^{-1}$ are the carrier lifetimes;$s_{1,2}$ are the gain saturation coefficients.\\
Now we shall demonstrate that depending on the laser systems' parameters eqs.(1-2) 
can allow for two regimes of lag synchronization between the the lasers' intensities  ( which are related to the electric field amplitudes by $I\propto\vert E \vert^{2}$).\\
\indent First we explore the possibility of perfect synchronization between the chaotic intensities of the master and slave lasers with the lag time equal to the coupling delay time - as is found in most experimental cases:
$$\hspace*{7cm}I_{1,\tau_{2}}=I_{2},\hspace*{7.8cm}(3)$$
(throughout this paper $x_{\tau}\equiv x(t-\tau)$).
We also assume an analogous synchronization manifold for the carrier densities:$N_{1,\tau_{2}}= N_{2}$. As was numerically shown in [8] such a synchronization manifold (with some modifications) can exist if $k_{1}=k_{2}$.
However as mentioned above, the modified synchronization manifold studied in [8] was not perfect even after introducing a constant scaling factor $a=1.016$ for the electric field amplitudes. In our notation the synchronization manifold considered in [8] would have been written as $a^{2}I_{1,\tau_{2}}=I_{2}$. We also note the following further difference between the synchronization manifolds considered in this work and those in paper [8]:namely above we assume the synchronization manifold  $N_{2}=N_{1,\tau_{2}}$, the analogous synchronization manifold in [8] is of the form $N_{2}=N_{1,\tau_{2}}+\Delta_{N}$, where $\Delta_{N}$ is some constant. We would like 
to emphasize that assumption of (even the slightest) structurally different forms of synchronization manifolds for the dynamical variables of the systems to be synchronized is rather unusual. At the same time we underline that even after such a modifications of the synchronizations manifolds made in [8], synchronization was not perfect. The authors of [8] with reference to C.R.Mirasso (Ref.[13] in [8]) indicate that perfect synchronization is possible if different photon lifetimes are assumed for the master and slave lasers. A similar idea for achieving perfect synchronization between master and slave lasers with coupling-delay lag time is indicated in our recent work [10].\\
In this paper we show that perfect synchronization can be achieved without the above 
mentioned modifications of the synchronization manifolds. Following [10], suppose that there are parameter mismatches between the master and slave laser photon decay rates:$\gamma_{1}\neq\gamma_{2}$. Using eq.(2) we write the dynamical equation for the $E_{\tau_{2}}$ in the following manner:
$$\hspace*{-1.7cm}\frac{dE_{1,\tau_{2}}}{dt}=\frac{(1+\imath\alpha_{1})}{2}(\frac{G_{1}(N_{1,\tau_{2}}-N_{01})}
{1+s_{1}\vert E_{1,\tau_{2}} \vert^{2}}-\gamma_{1})E_{1,\tau_{2}}+k_{1}E_{1,\tau_{1}+\tau_{2}}\exp(-\imath\omega\tau_{1}),$$
Assuming that the lasers parameters are identical (except for photon decay rates and feedback rates) we find that if $E_{1,\tau_{2}}$ and $E_{2}$ are related 
by $E_{1,\tau_{2}}=\pm E_{2}$ (consistent with the synchronization manifold (3)), then under the conditions
$$\hspace*{7cm}k_{1}=k_{2},\hspace*{7.8cm}(4)$$
and 
$$\hspace*{6cm}\frac{(1+\imath\alpha)}{2}\gamma_{1}=\frac{(1+\imath\alpha)}{2}\gamma_{2}\mp k_{3}\exp(-\imath\omega\tau_{2}),\hspace*{2.7cm}(5)$$
(where $\alpha=\alpha_{1}=\alpha_{2}$)
the equations for $E_{1,\tau_{2}}$ and $E_{2}$ become identical and the lag synchronization manifold (3) exists. We notice that the existence condition (4) of coupling-delay lag time synchronization (3) found here analytically is confirmed by numerical simulations in [8], where such a type of synchronization is demonstrated with identical feedback rates for the master and slave lasers: $k_{1}=k_{2}=20ns^{-1}$. But as was already mentioned above, in that paper it was also acknowledged that synchronization was not perfect. Our analytical approach demonstrates that for perfect synchronization in addition to condition (4), condition (5) is also required. Here we again recall that we use the generally accepted synchronization manifold:$I_{1,\tau_{2}}=I_{2}$ and $N_{1,\tau_{2}}=N_{2}$.\\
Thus in this paper we have demonstrated analytically for the first time that by taking into account difference in the photon lifetimes for the master and slave lasers one can obtain perfect synchronization. We have also established the relationship between the difference in the  photon lifetimes and the coupling rate between the lasers to achieve such perfect synchronization. The fact that most experimental work on the subject has reported synchronization with lag time of the the coupling delay time between the lasers is indicative of the fact that the lasers being synchronized had different cavity losses. \\
\indent Next we show that for identical lasers (except for the feedback rates $k_{1}$ and $k_{2}$) 
$$\hspace*{6cm}I_{1}=I_{2,\tau_{1}-\tau_{2}},\hspace*{8.2cm}(6)$$
is the lag synchronization manifold (we recall that for the lag synchronization $\tau_{2}>\tau_{1}$.)\\
We note that the synchronization manifold (6) is well-studied both numerically and analytically,see, e.g.[5] and we consider 
this case here only for completeness of our comparison with the results of [8]. For transparency of the analysis we again write the equation for $E_{2,\tau_{1}-\tau_{2}}$ 
explicitly:
$$\hspace*{0.2cm}\frac{dE_{2,\tau_{1}-\tau_{2}}}{dt}=\frac{(1+\imath\alpha_{2})}{2}(\frac{G_{2}(N_{2,\tau_{1}-\tau_{2}}-N_{02})}{1+s_{2}\vert E_{2,\tau_{1}-\tau_{2}} \vert^{2}}
-\gamma_{2})E_{2,\tau_{1}-\tau_{2}}+k_{2}E_{2,2\tau_{1}-\tau_{2}}\exp(-\imath\omega\tau_{1})+k_{3}E_{1,\tau_{1}}\exp(-\imath\omega\tau_{2}),$$
Now suppose that 
$E_{1}=E_{2,\tau_{1}-\tau_{2}}\exp(-\imath\omega(\tau_{1}-\tau_{2}))$, which is consistent with the synchronization manifold (6). Then it follows that under the condition 
$$\hspace*{6cm}k_{1}=k_{2}+k_{3},\hspace*{8.4cm}(7)$$
equations for $E_{1}$ and $E_{2,\tau_{1}-\tau_{2}}$ becomes 
identical (naturally we also assume that $N_{1}=N_{2,\tau_{1}-\tau_{2}}$) and perfect synchronization is possible. Thus under the condition (7) synchronization occur with the lag time $\tau_{1}-\tau_{2}$. This analytical result is also confirmed by numerical simulations in [8], where it was demonstrated that 
perfect synchronization when the synchronization error $E_{1}-E_{2,\tau_{1}-\tau_{2}}$ after short transient decays to zero is possible for the identical master and slave lasers with (in our notation) $k_{1}=20ns^{-1}, k_{2}=5ns^{-1}, k_{3}=15ns^{-1}$.\\
\indent To summarize, we have studied synchronization between unidirectionally coupled chaotic external cavity semiconductor lasers with two characteristic delay times, where the delay time in the coupling is  different from the delay time in the coupled 
systems themselves. We have demonstrated for the first time that parameter mismatches in photon decay rates for the master and slave lasers can explain the experimental observation that the lag time is equal to the coupling delay time and derived relevant 
existence conditions.\\
This work is supported by UK EPSRC under grants GR/R22568/01 and GR/N63093/01.\\

\end{document}